\documentclass[twocolumn,prb,aps,showpacs]{revtex4}
\usepackage{mathrsfs}
\usepackage{graphicx}
\begin{document}
\title {Exact solutions to the spin-2 Gross-Pitaevskii equations}
\author{Zhi-Hai Zhang}
\author{Yong-Kai Liu}
\author{Shi-Jie Yang\footnote{Corresponding author: yangshijie@tsinghua.org.cn}}
\affiliation{Department of Physics, Beijing Normal University, Beijing 100875, China}
\begin{abstract}
We present several exact solutions to the coupled nonlinear Gross-Pitaevskii equations which describe the motion of the one-dimensional spin-2 Bose-Einstein condensates. The nonlinear density-density interactions are decoupled by making use of the properties of Jacobian elliptical functions. The distinct time factors in each hyperfine state implies a "Lamor" procession in these solutions. Furthermore, exact time-evolving solutions to the time-dependent Gross-Pitaevskii equations are constructed through the spin-rotational symmetry of the Hamiltonian. The spin-polarizations and density distributions in the spin-space are analyzed.
\end{abstract}
 \maketitle

\section{Introduction}
The spinor Bose-Einstein condensates (BECs), which have been experimentally realized in optical potentials, exhibit a rich variety of magnetic effects. Due to the internal degrees of freedom, they give rise to phenomena that are not present in the single component BECs\cite{Stenger,Miesner,Kobayashi}, such as magnetic crystallization, spin textures, and fractional vortices, ect. In a spinor condensate, there exists an interplay between superfluidity and magnetism associated to the spin-gauge symmetry in the Hamiltonian. A direct consequence is that the ferromagnetic BECs spontaneously induce a supercurrent as the spin is locally rotated\cite{Ueda,Kawaguchi}.

The mean-field motion of the dilute spinor condensates is governed by the coupled Gross-Pitaevskii equations (GPEs)\cite{Gross,Ginzburg,Dalfovo,Coen,N.Z.}. There are many theoretical works that numerically solve GPEs or the corresponding stationary equations\cite{Wang,Gammal,Cerimele,Sun}. Analytical solutions to the GPEs are generally difficult because of the nonlinear density-density coupling between the atoms as well as the spin-spin coupling between the hyperfine states. Many efforts have been contributed to the one-dimensional (1D) soliton limits\cite{Khawaja,Nistazakis}, mainly for the $F=1$ condensates. Nevertheless, exact analytical solutions for the spinor BECs, especially for the $F=2$ condensates, are absent in literature. In a previous work, we have presented exact solutions to the $F=1$ GPEs\cite{Zhang}. In this paper, we construct exact non-solitary solutions to the $F=2$ GPEs in 1D and give more insights into the internal structure of the states. The solutions are of complex form and are expressed in combinations of the Jacobian elliptical functions.

The paper is organized as follows. In Sec.II we describe the 1D coupled nonlinear GPEs for $F=2$ BECs. A set of generalized stationary equations are deduced. In Sec.III we present the special forms of analytical solutions in which only two of the components are nonzero. In Sec.IV, the associated time-evolving solutions to the special solutions are constructed by making using of the spin-rotational symmetry of the Hamiltonian. Section V includes a brief summary.

\section{Equations of motion}
The spinor condensate formed by spin-2 atoms is described by a macroscopic wave function with five hyperfine states $\Psi=(\psi_{+2},\psi_{+1},\psi_0,\psi_{-1},\psi_{-2})^T$. The mean-field Hamiltonian is expressed as\cite{Ciobanu,Ho}
\begin{eqnarray}
H=\int d\textbf{r}&&\{\sum_{m=-2}^2\psi_m^*[\frac{-\hbar^2}{2M}\nabla^2+V(\textbf{r})]\psi_m+\nonumber\\
&&\frac{\bar{c}_0}{2}n^2+\frac{\bar{c}_1}{2}|\textbf{F}|^2+\frac{\bar{c}_2}{2}|A_{00}|^2\},\label{Hamiltonian}
\end{eqnarray}
where $\textbf{F}=\psi_m^*\hat{\textbf{F}}_{mn}\psi_n$ with $\hat{\textbf{F}}^i (i=x,y,z)$ the
$5\times5$ spin matrices. The coupling constants $\bar{c}_0$, $\bar{c}_1$ and $\bar{c}_2$ are related to scattering lengths $a_0$, $a_2$ and $a_4$ of the two colliding bosons, with total angular momenta $0$, $2$ and $4$, by $\bar{c}_0=4\pi\hbar^2(a_4-a_2)/7M$, $\bar{c}_1=4\pi\hbar^2(3a_4+4a_2)/7M$,
$\bar{c}_2=4\pi\hbar^2(3a_4-10a_2+7a_0)/7M$. The total atom density is $n=|\psi_2|^2+|\psi_1|^2+|\psi_0|^2+|\psi_{-1}|^2+|\psi_{-2}|^2$.
The amplitude of the spin-singlet pair $A_{00}=(2\psi_2\psi_{-2}-2\psi_1\psi_{-1}+\psi_0^2)/\sqrt{5}$. $V(\textbf{r})$ is the external potential. Hamiltonian (\ref{Hamiltonian}) possesses the $U(1)_{\textrm{phase}}\times SO(3)_{\textrm{spin}}$ symmetry. The energies are degenerate for an arbitrary state $\Psi$ and its globally spin-rotational states $\Psi^\prime=U \Psi$, where $U$ is the $5\times5$ rotational matrix in the spin space which is expressed by the Euler angles as $U(\alpha,\beta,\gamma)=e^{-i\hat{F}_z\alpha}e^{-i\hat{F}_y\beta}e^{-i\hat{F}_z\gamma}$. In the ground state,  the symmetry is spontaneously broken in several different ways, leading to a number of possible phases\cite{Ho,chang,Murata,Imambekov}.

We are concerned with the quasi-1D $F=2$ BECs in a uniform external potential
$(V(\textbf{r})=0)$. The dynamical motion of the spinor wave functions are governed by
$i\partial_t\psi_m=\delta H/\delta\psi_m^*$, which are explicitly written as the coupled nonlinear GPEs,
\begin{eqnarray}
i\hbar\frac{\partial \psi_{\pm2}}{\partial t}&=&[-\frac{\hbar^2\nabla^2}{2M}+c_0 n \pm 2c_1 F_z]\psi_{\pm2}+c_1F_{\mp} \psi_{\pm1}\nonumber\\
&&+\frac{c_2}{\sqrt{5}}A\psi_{\mp2}^*\nonumber\\
i\hbar\frac{\partial \psi_{\pm1}}{\partial t}&=&[-\frac{\hbar^2\nabla^2}{2M}+c_0 n \pm c_1 F_z]\psi_{\pm1}\nonumber\\
&&+c_1(\frac{\sqrt{6}}{2}F_{\mp}\psi_0+F_{\pm}\psi_{\pm2})-\frac{c_2}{\sqrt{5}}A\psi_{\mp1}^*\nonumber\\
i\hbar\frac{\partial \psi_{0}}{\partial t}&=&[-\frac{\hbar^2\nabla^2}{2M}+c_0 n]\psi_0+\frac{\sqrt{6}}{2}c_1(F_+\psi_1+F_-\psi_{-1})\nonumber\\
&&+\frac{c_2}{\sqrt{5}}A\psi_0^*\nonumber\\\label{GPE}
\end{eqnarray}
where $F_+=F_-^*=2(\psi_2^*\psi_1+\psi_{-1}^*\psi_{-2})+\sqrt{6}(\psi_1^*\psi_0+\psi_0^*\psi_{-1})$ and
$F_z=2(|\psi_2|^2-|\psi_{-2}^2)+|\psi_1|^2-|\psi_{-1}|^2$. $c_0=\bar{c}_0/2a_\perp^2$,
$c_1=\bar{c}_1/2a_\perp^2$ and $c_2=\bar{c}_2/2a_\perp^2$ are the reduced coupling constants with
$a_\perp$ the transverse width of the quasi-1D system.

Below we choose $\hbar=M=1$ as the units for convenience. $c_0$, $c_1$ and $c_2$ are treated as free
parameters. By substituting the wavefunction $\Psi(x,t)$ with
\begin{equation}
\left(
          \begin{array}{c}
          \psi_2(x,t) \\
          \psi_1(x,t) \\
          \psi_0(x,t) \\
          \psi_{-1}(x,t) \\
          \psi_{-2}(x,t) \\
          \end{array}
        \right)
\rightarrow\left(
          \begin{array}{c}
            \psi_2(x)e^{-i(\mu+\mu_2)t} \\
            \psi_1(x)e^{-i(\mu-\mu_2)t}\\
            \psi_0(x)e^{-i\mu t} \\
            \psi_{-1}(x)e^{-i(\mu-\mu_1)t}\\
            \psi_{-2}(x)e^{-i(\mu+\mu_1)t} \\
          \end{array}
        \right),
\end{equation}
we obtain the generalized stationary GPEs as,
\begin{eqnarray}
(\mu \pm \mu_2) \psi_{\pm2}&=&[-\frac{1}{2}\partial_x^2+c_0 n \pm 2c_1 F_z]\psi_{\pm2}+c_1F_{\mp} \psi_{\pm1}\nonumber\\
&&+\frac{c_2}{\sqrt{5}}A\psi_{\mp2}^*\nonumber\\
(\mu \pm \mu_1) \psi_{\pm1}&=&[-\frac{1}{2}\partial_x^2+c_0 n \pm c_1 F_z]\psi_{\pm1}\nonumber\\
&&+c_1(\frac{\sqrt{6}}{2}F_{\mp}\psi_0+F_{\pm}\psi_{\pm2})-\frac{c_2}{\sqrt{5}}A\psi_{\mp1}^*\nonumber\\
\mu \psi_0&=&[-\frac{1}{2}\partial_x^2+c_0 n]\psi_0+\frac{\sqrt{6}}{2}c_1(F_+\psi_1+F_-\psi_{-1})\nonumber\\
&&+\frac{c_2}{\sqrt{5}}A\psi_0^*.\nonumber\\
\label{stationary}
\end{eqnarray}
It is notable that the solutions to the Eq.(\ref{stationary}) are not strictly "stationary" states since each component contains a distinct time-dependent phase factor, given $\mu_1$ and $\mu_2$ are not equal to zero. As we shall show, a "Lamor" procession in the spin space is associated to these states. The parameters $\mu_1$ and $\mu_2$ play the roles of linear Zeeman energies. Nevertheless, the density distribution of each hyperfine state are time-invariant. Hence we simply call these states as stationary states.

The periodical boundary conditions
\begin{equation}
\psi_m(1)=\psi_m(0),\hspace{3mm} \psi_m^\prime(1)=\psi_m^\prime(0),\label{periodic}
\end{equation}
is adopted. In the following we consider two types of complex solutions to the stationary equations (\ref{stationary}).

\section{Special solutions}
In order to decouple the nonlinear spin-spin interactions in Eq.(\ref{stationary}), we consider the special cases with only two of the hyperfine states are nonzero. The nonlinear density-density interactions are decoupled by making use of the unique properties of the Jacobian elliptical functions, as shown in the follows. The general forms of solutions in which all components are nonzero are obtained by the applying a rotation in the spin space, which will be addressed in the next section.
\subsection{Type A solution}
We first take the following ansatz as the solution to the nonlinear Eqs.(\ref{stationary}),
\begin{equation}
\left(
          \begin{array}{c}
          \psi_2(x) \\
            \psi_1(x) \\
            \psi_0(x) \\
            \psi_{-1}(x) \\
             \psi_{-2}(x) \\
          \end{array}
        \right)
=\left(
          \begin{array}{c}
            f(x)e^{i\theta(x)} \\
            0\\
            0 \\
            D\textrm{sn}(kx,m)\\
            0 \\
          \end{array}
        \right),\label{Type A}
\end{equation}
where $f(x)=\sqrt{A+B\textrm{cn}^2(kx,m)}$ and $A$, $B$ and $D$ are real constants. \textrm{sn} and
\textrm{cn} are the Jacobian elliptical functions and $k=4jK(m)$ with $m$ the modulus $(0<m<1)$. In the
context we always take the number of periods $j=2$ as examples. One has
\begin{equation}
\textrm{sn}^2=\frac{|\psi_{-1}|^2}{D^2},\\
\textrm{cn}^2=\frac{|\psi_2|^2-A}{B}.
\end{equation}
By substituting the relations into the coupled GPEs(\ref{stationary}) and making use of the relations between the Jacobian elliptical functions, one obtain the decoupled equations
\begin{eqnarray}
\tilde{\mu}_{2}\psi_{2}&=&-\frac{1}{2}\psi_2^{\prime\prime}+\tilde{\gamma}_2|\psi_2|^2\psi_2,\nonumber\\
\tilde{\mu}_{-1}\psi_{-1}&=&-\frac{1}{2}\psi_{-1}^{\prime\prime}+\tilde{\gamma}_{-1}|\psi_{-1}|^2\psi_{-1},
\label{decouple A}
\end{eqnarray}
where the effective chemical potentials $\tilde{\mu}_m$ and effective nonlinear coupling constants $\tilde{\gamma}_m$ are respectively defined as
\begin{eqnarray}
\left\{\begin{array}{ll}\tilde{\mu}_2=\mu+\mu_2-c_0(A+D^2)+c_0\frac{A}{B}(B-D^2)\\
\ \ \ \ \ \ -2c_1(2A-D^2)+2c_1\frac{A}{B}(2B+D^2)\\
\tilde{\gamma}_2=\frac{c_0}{B}(B-D^2)+\frac{2c_1}{B}(2B+D^2),
\end{array}\right.
\end{eqnarray}
and
\begin{equation}
\left\{\begin{array}{ll}\tilde{\mu}_{-1}=\mu-\mu_1-c_0(A+B)+2c_1(A+B)\\
\tilde{\gamma}_{-1}=\frac{c_0}{D^2}(D^2-B)+\frac{c_1}{D^2}(2B+D^2).
\end{array}\right.
\end{equation}

The decoupled Eqs.(\ref{decouple A}) can be self-consistently solved to yield,
\begin{eqnarray}
\tilde{\mu}_{-1}=\frac{1}{2}k^2(1+m^2),\hspace{3mm} \tilde{\gamma}_{-1}=\frac{m^2k^2}{D^2}.
\label{A1}
\end{eqnarray}
\begin{eqnarray}
B=-\frac{m^2k^2}{\tilde{\gamma}_2},\hspace{3mm} A=\frac{2\tilde{\mu}_2-k^2(1-2m^2)}{3\tilde{\gamma}_2}.
\label{A2}
\end{eqnarray}

The phase in Eq.(\ref{Type A}) is calculated as
\begin{equation}
\theta(x)=\int_0^x\frac{\alpha_1}{f^2(\xi)}d\xi,
\end{equation}
where $\alpha_1=\pm(2\tilde{\mu}_2A^2-2\tilde{\gamma}_2A^3+k^2(1-m^2)AB)^{\frac{1}{2}}$ is the integral
constant. The periodical boundary conditions (\ref{periodic}) require that the amplitudes and phase
satisfy
\begin{equation}
f(1)=f(0),\ \ \ \theta(1)-\theta(0)=2j\pi\times n, \label{boundaryA}
\end{equation}
respectively, where $n$ is an integer. The periodical condition for the phase is satisfied by properly adjusting the modulus $m$ of the Jacobian elliptical functions. In the calculations, we fix the modulus $m$ and $\mu_1$, $\mu_2$ while adjust $\mu$ to fulfill the periodical boundary conditions (\ref{boundaryA}).

\begin{figure}
\begin{center}
\includegraphics*[width=10cm]{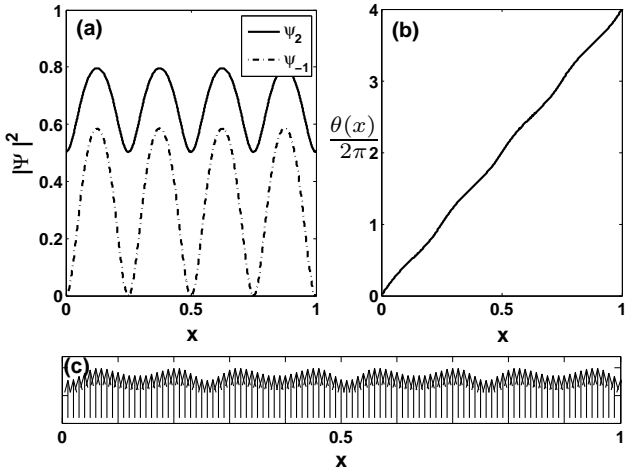}
\caption{The density (a) and the phase (b) distributions for
solution (\ref{Type A}) with $n=2$, $m=0.81$. The physical
parameters are $c_0=8$, $c_1=7$, $\mu=201.9003$, $\mu_1=60$, and
$\mu_2=600$. (c) The distribution of the spin-polarization.}
\end{center}
\end{figure}

Figure 1 display the distributions of the density, the phase and the spin-polarization of the solution (\ref{Type A}). The relevant parameters are $c_0=8$, $c_1=7$, $m=0.81$, $\mu=201.9003$, $\mu_1=60$, $\mu_2=600$, and $n=2$. Numerically, we get $\tilde\mu_2=655.2336$ and $\tilde\gamma_2=24$. It implies that the effective interactions in the $m=-1$ hyperfine state should be repulsive while in the $m=+2$ hyperfine state should be repulsive. One notes that $F_x(x)=F_y(x)\equiv 0$ and $A_{00}(x)=0$. Obviously, the spin-polarization vector ${\bf F}(x)$ is not enough to exhibit the spin configurations for the $F=2$ condensates. In order to reveal the "Lamor" precession of the spin, in Fig.2 we display the density distribution in the spin space. At the nodes of the $\psi_{-1}$ ($x=0$, $0.5$, and $1.0$) it becomes fully spin-polarized with the spin-configurations exhibit the axial symmetry. At other spatial positions the spin-configuration exhibit the tetrahedral symmetry, which subsequently rotate an angle to the adjacent positions. As we have stressed, although state (\ref{Type A}) is the solution to the stationary equation (\ref{stationary}), it contains a spin procession with frequency $\omega=\mu_2+\mu_1$. Fig.2 snapshots this procession at $t=0$, $t=T/8$, and $t=T/3$, where period $T=2\pi/\omega$.

\begin{figure}
\begin{center}
\includegraphics*[width=10cm]{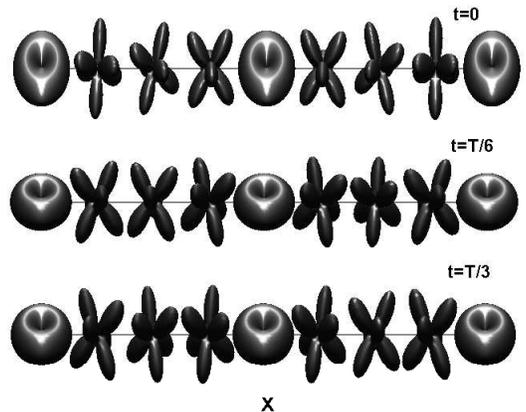}
\caption{Snapshots of the density distribution for (\ref{Type A}) (attached by the time factor) in the spin space at $t=0$, $t=T/8$, and $t=T/3$, respectively. The procession period is $T=2\pi/\omega$.}
\end{center}
\end{figure}

\subsection{Type B solution}
We next take an alternative ansatz as the solution to the stationary Eqs.(\ref{stationary}),
\begin{equation}
\left(
          \begin{array}{c}
          \psi_2(x) \\
            \psi_1(x) \\
            \psi_0(x) \\
            \psi_{-1}(x) \\
             \psi_{-2}(x) \\
          \end{array}
        \right)
=\left(
          \begin{array}{c}
            f(x)e^{i\theta(x)} \\
            0\\
            0 \\
            0\\
            D\textrm{cn}(kx,m)\\
          \end{array}
        \right),\label{Type B}
\end{equation}
where $f(x)=\sqrt{A+B\textrm{sn}^2(kx,m)}$ and $A$, $B$ and $D$ are real constants. One has
\begin{equation}
\textrm{cn}^2=\frac{|\psi_{-2}|^2}{D^2},\hspace{3mm}
\textrm{sn}^2=\frac{|\psi_2|^2-A}{B}.
\end{equation}

By the same way, we obtain the decoupled equations as
\begin{eqnarray}
\tilde{\mu}_{2}\psi_{2}&=&-\frac{1}{2}\psi_2^{\prime\prime}+\tilde{\gamma}_2|\psi_2|^2\psi_2,\nonumber\\
\tilde{\mu}_{-2}\psi_{-2}&=&-\frac{1}{2}\psi_{-2}^{\prime\prime}+\tilde{\gamma}_{-2}|\psi_{-2}|^2\psi_{-2},
\label{decouple B}
\end{eqnarray}
where the effective chemical potentials $\tilde{\mu}_m$ and effective nonlinear coupling constants $\tilde{\gamma}_m$ are respectively defined as
\begin{eqnarray}
\left\{\begin{array}{ll}\tilde{\mu}_2=\mu+\mu_1-c_0(A+D^2)+c_0\frac{A}{B}(B-D^2)\\
\ \ \ \ -4c_1(A-D^2)+4c_1\frac{A}{B}(B+D^2)-\frac{2c_2D^2}{5B}(A+B)\\
\tilde{\gamma}_2=c_0(1-\frac{D^2}{B})+4c_1(1+\frac{D^2}{B})-\frac{2c_2D^2}{5B},
\end{array}\right.
\end{eqnarray}
and
\begin{equation}
\left\{\begin{array}{ll}\tilde{\mu}_{-2}=\mu+\mu_2-c_0(A+B)+4c_1(A+B)-\frac{2c_2}{5}(A+B)\\
\tilde{\gamma}_{-2}=c_0(1-\frac{B}{D^2})+4c_1(1+\frac{B}{D^2})-\frac{2c_2B}{5D^2}.
\end{array}\right.
\end{equation}
The decoupled Eqs.(\ref{decouple B}) are self-consistently solved to yield
\begin{eqnarray}
\tilde{\mu}_{-2}=\frac{1}{2}k^2(1-2m^2),\hspace{3mm}\tilde{\gamma}_{-2}=-\frac{m^2k^2}{D^2},
\label{B1}
\end{eqnarray}
\begin{eqnarray}
B=\frac{m^2k^2}{\tilde{\gamma}_2},\hspace{3mm} A=\frac{2\tilde{\mu}_2-k^2(1+m^2)}{3\tilde{\gamma}_2}.\label{B2}
\end{eqnarray}

The phase in Eq.(\ref{Type B}) is
\begin{equation}
\theta(x)=\int_0^x\frac{\alpha_2}{f^2(\xi)}d\xi,
\end{equation}
where $\alpha_2=\pm(2\tilde{\mu}_2A^2-2\tilde{\gamma}_2A^3+k^2AB)^{\frac{1}{2}}$ is the integral constant. The amplitudes and phase should satisfy the periodical boundary conditions (\ref{boundaryA}) and the modulus $m$ is accordingly determined.
\begin{figure}
\begin{center}
\includegraphics*[width=10cm]{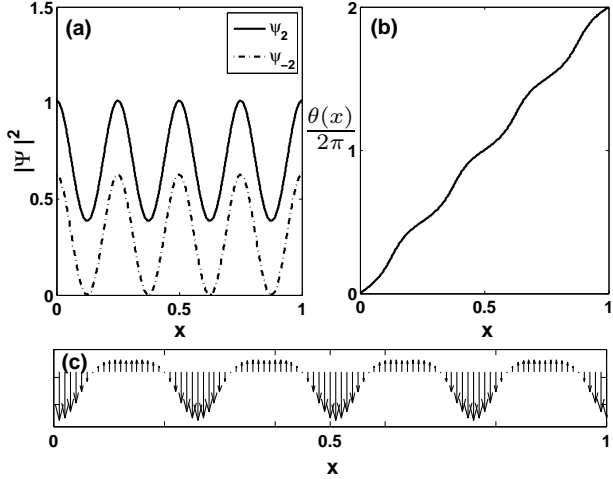}
\caption{The density (a) and the phase (b) distributions for
solution (\ref{Type B}) with $n=1$, $m=0.4$. The physical parameters
are $c_0=-100$, $c_1=-80$, $c_2=-20$, $\mu=55.8083$, $\mu_1=-40$,
and $\mu_2=20$. (c) The distribution of the spin-polarization.}
\end{center}
\end{figure}

Figure 3 display the distributions of the density, the phase, and the spin-polarization of the solution
(\ref{Type B}). The physical parameters are chosen as $c_0=-100$, $c_1=-80$, $c_2=-20$, $\mu=55.8083$,
$\mu_1=-40$, and $\mu_2=20$, $n=1$ $m=0.4$.  Numerically, we get $\tilde\mu_2=33.0909$ and $\tilde\gamma_2=-208$. Analogously, $F_x(x)=F_y(x)\equiv 0$ and $A_{00}(x)\neq 0$. In order to exhibit the spin procession of the state, we show in Fig.4 the density distribution in the spin space at $t=0$, $t=T/8$, and $t=T/3$, respectively. The procession frequency is $\omega=\mu_2-\mu_1$ and $T=2\pi/\omega$. The spin is fully polarized as the nodes of hyperfine state $\psi_{-2}$. At other positions, the density distribution has $C_{4z}$ symmetry and is obviously distinct to that of solution (\ref{Type A}).
\begin{figure}
\begin{center}
\includegraphics*[width=10cm]{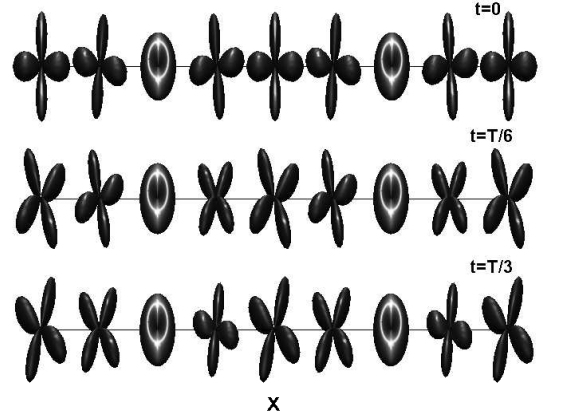}
\caption{Snapshots of the density distribution for (\ref{Type B}) in the spin space at $t=0$, $t=T/8$, and $t=T/3$, respectively. The procession period is $T=2\pi/\omega$.}
\end{center}
\end{figure}

\section{General solutions}
The more general time-evolving solutions to the GPEs (\ref{GPE}) can be obtained by making use of the symmetry of the Hamiltonian. If $\Psi(x,t)$ is a solution to the GPEs (\ref{GPE}), then $\Psi^\prime(x,t)=U \Psi(x,t)$ is also a solution, where $U$ is an arbitrary rotational transformation in the spin space which is explicitly written with the Euler angles as\cite{Makela}
\begin{widetext}
\begin{equation}
U=      \left(
          \begin{array}{ccccc}
            e^{-2i(\alpha+\gamma)}\cos^4\frac{\beta}{2} & -e^{-i(2\alpha+\gamma)}\sin\beta\cos^2\frac{\beta}{2} & \frac{\sqrt{6}}{4}e^{-2i\alpha}\sin^2\beta & -e^{-i(2\alpha-\gamma)}\sin\beta\sin^2\frac{\beta}{2} & e^{-2i(\alpha-\gamma)}\sin^4\frac{\beta}{2} \\
            e^{-i(\alpha+2\gamma)}\sin\beta\cos^2\frac{\beta}{2} & \frac{1}{2}e^{-i(\alpha+\gamma)}(\cos\beta+\cos2\beta) & -\frac{\sqrt{6}}{4}e^{-i\alpha}\sin2\beta & \frac{1}{2}e^{-i(\alpha-\gamma)}(\cos\beta-\cos2\beta) & -e^{-i(\alpha-2\gamma)}\sin\beta\sin^2\frac{\beta}{2}\\
            \frac{\sqrt{6}}{4}e^{-2i\gamma}\sin^2\beta & \frac{\sqrt{6}}{4}e^{-i\gamma}\sin2\beta & \frac{1}{4}(1+3\cos2\beta) & -\frac{\sqrt{6}}{4}e^{i\gamma}\sin2\beta & \frac{\sqrt{6}}{4}e^{2i\gamma}\sin^2\beta \\
            e^{i(\alpha-2\gamma)}\sin\beta\sin^2\frac{\beta}{2} & \frac{1}{2}e^{i(\alpha-\gamma)}(\cos\beta-\cos2\beta) & \frac{\sqrt{6}}{4}e^{i\alpha}\sin2\beta & \frac{1}{2}e^{i(\alpha+\gamma)}(\cos\beta+\cos2\beta) & -e^{i(\alpha+2\gamma)}\sin\beta\cos^2\frac{\beta}{2} \\
            e^{2i(\alpha-\gamma)}\sin^4\frac{\beta}{2} & e^{i(2\alpha-\gamma)}\sin\beta\sin^2\frac{\beta}{2} & \frac{\sqrt{6}}{4}e^{2i\alpha}\sin^2\beta & e^{i(2\alpha+\gamma)}\sin\beta\cos^2\frac{\beta}{2} & e^{2i(\alpha+\gamma)}\cos^4\frac{\beta}{2} \\
          \end{array}
        \right)
\end{equation}
\end{widetext}
In the following we fix the three Euler angles as $\alpha=7\pi/13$, $\beta=\pi/4$ and $\gamma=\pi/11$.

\subsection{Solution associated to Type A}
By attaching the time factors to the solution (\ref{Type A}) and then applying the rotational transformation, we obtain the time-evolving solutions to Eqs.(\ref{GPE}),
\begin{widetext}
\begin{eqnarray}
\psi^\prime= e^{-i\mu t}\left(
          \begin{array}{c}
            \phi_2\cos^4\frac{\beta}{2}e^{-2i(\alpha+\gamma)}e^{-i\mu_2t}-\phi_{-1}\sin^2\frac{\beta}{2}\sin\beta e^{-i(2\alpha-\gamma)}e^{i\mu_1t} \\
            \phi_2\cos^2\frac{\beta}{2}\sin\beta e^{-i(\alpha+2\gamma)}e^{-i\mu_2t}-\frac{1}{2}\phi_{-1}(\cos2\beta-\cos\beta)e^{-i(\alpha-\gamma)}e^{i\mu_1t}\\
            \frac{\sqrt{6}}{4}\phi_2\sin^2\beta e^{-2i\gamma}e^{-i\mu_2t}-\frac{\sqrt{6}}{4}\phi_{-1}\sin2\beta e^{i\gamma}e^{i\mu_1t}\\
            \phi_2\sin^2\frac{\beta}{2}\sin\beta e^{i(\alpha-2\gamma)}e^{-i\mu_2 t}+\frac{1}{2}\phi_{-1}(\cos2\beta+\cos\beta)e^{i(\alpha+\gamma)}e^{i\mu_1t}\\
            \phi_2\sin^4\frac{\beta}{2}e^{2i(\alpha-\gamma)}e^{-i\mu_2t}+\phi_{-1}\sin\beta\cos^2\frac{\beta}{2}e^{i(2\alpha+\gamma)}e^{i\mu_1t}\\
          \end{array}\label{non-A}
        \right),
\end{eqnarray}
\end{widetext}
where $\phi_2=\sqrt{A+B\textrm{cn}^2(kx,m)}e^{i\theta(x)}$ and $\phi_{-1}=D\textrm{sn}(kx,m)$. It should be noted that even at $t=0$, the solution (\ref{non-A}) is no longer a solution to the stationary Eq.(\ref{stationary}). The temporal evolution of the density distribution of each component is shown in Fig.5. The rightmost column of Fig.5 displays the snapshots of the spin rotation at $x=0.143$.

\begin{figure}
\begin{center}
\includegraphics*[width=9cm]{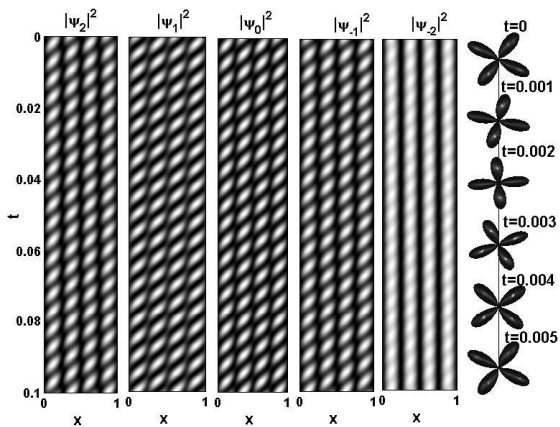}
\caption{Temporal evolution of the density of solution (\ref{non-A}). The parameters are the same as in Fig.1. The rightmost column displays the spin rotation at $x=0.143$.}
\end{center}
\end{figure}

\subsection{Solution associated to Type B}
Similarly, we obtain the general time-evolving solution associated to (\ref{Type B}) as,
\begin{widetext}
\begin{eqnarray}
\psi^\prime= e^{-i\mu t}\left(
          \begin{array}{c}
            \phi_2\cos^4\frac{\beta}{2}e^{-2i(\alpha+\gamma)}e^{-i\mu_2t}+\phi_{-2}\sin^4\frac{\beta}{2}e^{-2i(\alpha-\gamma)}e^{i\mu_1t}\\
            \phi_2\cos^2\frac{\beta}{2}\sin\beta e^{-i(\alpha+2\gamma)}e^{-i\mu_2t}-\phi_{-2}\sin^2\frac{\beta}{2}\sin\beta e^{-i(\alpha-2\gamma)}e^{i\mu_1t}\\
            \frac{\sqrt{6}}{4}\phi_2\sin^2\beta e^{-2i\gamma}e^{-i\mu_2t}+\frac{\sqrt{6}}{4}\phi_{-2}\sin^2\beta e^{2i\gamma}e^{i\mu_1t}\\
            \phi_2\sin^2\frac{\beta}{2}\sin\beta e^{i(\alpha-2\gamma)}e^{-i\mu_2t}-\phi_{-2}\cos^2\frac{\beta}{2}\sin\beta e^{i(\alpha+2\gamma)}e^{i\mu_1t}\\
            \phi_2\sin^4\frac{\beta}{2}e^{2i(\alpha-\gamma)}e^{-i\mu_2t}+\phi_{-2}\cos^4\frac{\beta}{2}e^{2i(\alpha+\gamma)}e^{i\mu_1t}\\
          \end{array}\label{non-B}
        \right),
\end{eqnarray}
\end{widetext}
where $\phi_1=\sqrt{A+B\textrm{cn}^2(kx,m)}e^{i\theta(x)}$ and $\phi_{-2}=D\textrm{sn}(kx,m)$. Figure 6 show the temporal evolution of the density of solution (\ref{non-B}). The rightmost column is the snapshots of the spin rotation at $x=0.2$.

\begin{figure}
\begin{center}
\includegraphics*[width=9cm]{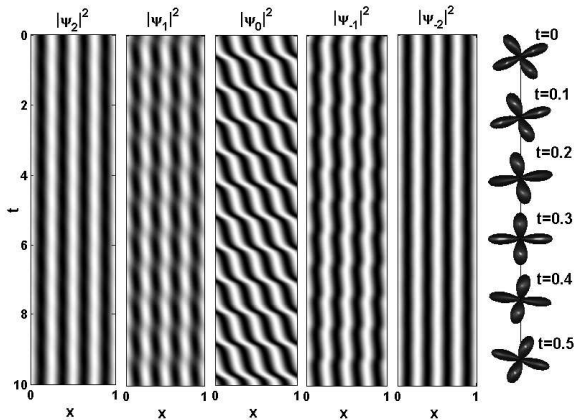}
\caption{Temporal evolution of the density of solution (\ref{non-B}). The parameters are the
same as in Fig.3. The rightmost column displays the spin rotation at $x=0.2$.}
\end{center}
\end{figure}

\section{Summary}
In summary, we have presented two classes of analytical stationary solutions to the 1D coupled
nonlinear GPEs which govern the dynamics of the spin-2 condensates. Obviously, we can obtain a lot of
other exact solutions with different combinations of the Jacobian elliptical functions. The exact time-evolving solutions are also constructed. The spin-polarization and the spin procession are addressed.

This work is supported by the funds from the Ministry of Science and Technology of China under Grant
No.2012CB821403.

\end{document}